\documentclass[a4paper,11pt]{article}
\usepackage{pos}
\usepackage{multicol}
\usepackage{setspace}

\title{A New Search for Neutrino Point Sources with IceCube}
 \ShortTitle{A New Search for Neutrino Point Sources with IceCube}

\author{The IceCube Collaboration \\{\normalsize \normalfont(a complete list of authors can be found at the end of the proceedings)}}

\emailAdd{hans.niederhausen@icecube.wisc.edu}
\emailAdd{theo.glauch@tum.de}

\abstract{The IceCube Neutrino Observatory, deployed inside the deep glacial ice at the South Pole, is the largest neutrino telescope in the world. While eight years have passed since IceCube discovered a diffuse flux of high-energy astrophysical neutrinos, the sources of the vast majority of these neutrinos remain unknown. Here, we present a new search for neutrino point sources that improves the accuracy of the statistical analysis, especially in the low energy regime. We replaced the usual Gaussian approximations of IceCube's point spread function with precise numerical representations, obtained from simulations, and combined them with new machine learning-based estimates of event energies and angular errors. Depending on the source properties, the new analysis provides improved source localization, flux characterization and thereby discovery potential (by up to 30\%) over previous works. The analysis will be applied to IceCube data that has been recorded with the full 86-string detector configuration from 2011 to 2020 and includes improved detector calibration.

\vspace{4mm}
{\bfseries Corresponding authors:}
Chiara Bellenghi$^{1}$, Theo Glauch$^{1}$, Christian Haack$^{1}$, Tomas Kontrimas$^{1}$, Hans Niederhausen$^{*1,3}$, Rene Reimann$^{2}$, Martin Wolf$^{1}$\\[4mm]
{$^{1}$ \itshape Department of Physics, Technical University of Munich, James-Franck-Straße 1, 85748 Garching bei München, Germany}\\
{$^{2}$ \itshape Institute of Physics - QUANTUM, Johannes Gutenberg University, Staudinger Weg 7, 55099 Mainz, Germany}\\
{$^{3}$ \itshape Department of Physics and Astronomy, Michigan State University, Biomedical and Physical Sciences, 567 Wilson Road, East Lansing, MI 48824, USA}\\[4mm]
$^*$ Presenter

\FullConference{37$^{\rm{th}}$ International Cosmic Ray Conference (ICRC 2021)\\
		July 12th -- 23rd, 2021\\
		Online -- Berlin, Germany}

}

\begin{document}
\maketitle

\noindent\textbf{Introduction}

\noindent IceCube has been measuring a diffuse flux of high-energy astrophysical neutrinos with increasing precision and significance since 2013 \cite{Aartsen:2013jdh, Aartsen:2020aqd, Stettner:2019tok}. The flux appears to be isotropic, to have equal contributions from all neutrino flavors and to follow a power-law with a spectral index of $\gamma \approx 2.5$ in the TeV to PeV energy range. In 2017, IceCube identified the gamma-ray flaring blazar TXS 0506+056 as the first compelling, potentially time-variable source of extra-galactic neutrino emission \cite{IceCube:2018dnn, IceCube:2018cha}. Yet, the population(s) of sources responsible for the vast majority of this diffuse flux remain unknown and progress in searches for neutrino point sources is needed.
IceCube has performed several searches for time-integrated neutrino point sources. The most recent analysis reported a 2.9 $\sigma$ excess of soft-spectrum ($\gamma \approx 3.2$) neutrino events from the direction of the active galaxy NGC 1068 \cite{Aartsen:2019fau}. All of these searches were based on the un-binned likelihood formalism presented in \cite{Braun:2008bg}, which exploits spatial clustering as well as energy information to discriminate a point-like astrophysical neutrino signal from the background of atmospheric and diffuse astrophysical neutrinos. While this method has been quite successful, it is limited by approximations in the corresponding likelihood function. We have therefore revised the likelihood from first principles and used new numerical techniques to extract a more accurate description of the underlying probability density functions (pdfs) from Monte-Carlo (MC) simulations. Here, we benefited greatly from significant advances in the understanding and hence modeling of the IceCube detector that were made throughout recent years.  Furthermore, new reconstruction algorithms for the energy and the angular uncertainty of each event were developed. Overall, these new methods improve the probability to detect new astrophysical neutrino sources and, in such cases, also the measurement of their properties.   \\

\noindent\textbf{Event Selection and Data Sample} 

\noindent Depending on the flavor of the neutrino and the type of its interaction in or near the IceCube detector different secondary particles are produced. Most relevant for neutrino point source searches are muons, produced in charged-current $\nu_{\mu}$ interactions with the nucleons in the ice, that travel for several kilometers. Such track events can reach angular resolutions below 0.4 degrees above $\sim 100\,\mathrm{TeV}$. Moreover, these long tracks effectively increase the detector volume as the neutrino interaction point can lie far outside the instrumented volume. Track events have also been used to measure the diffuse astrophysical neutrino flux in the Northern Hemisphere \cite{Stettner:2019tok}, where the Earth shields the detector from the overwhelming background of atmospheric muons. The analysis presented here uses the same event selection criteria that achieve a neutrino purity of $\sim 99.7\%$. The sample consists of well-reconstructed tracks in the energy range between $100\,\mathrm{GeV}$ and several PeV, and has an event rate of $\sim 2.5\,\mathrm{mHz}$, strongly dominated by atmospheric neutrinos. Based on the diffuse flux measured in \cite{Stettner:2019tok} we expect the rate of astrophysical neutrinos to be $\sim 0.02\,\mathrm{mHz}$, corresponding to a selection efficiency of $\sim 99\%$. Scaling up to 9 years of data taking with the complete detector, this results in a data sample of around 670,000 events of which $\mathcal{O}(1000)$ are expected to be of cosmic origin. This highly-pure event selection shows good agreement between the experimental data and the MC simulations. The sample is therefore well-suited to be analyzed using the new point source analysis methods presented here, which rely more heavily on simulated data for the modelling of the likelihood function and the calculation of the corresponding test-statistics distributions than previous works.\\

\noindent\textbf{Constructing a More Accurate Likelihood Function}

\noindent Maximum likelihood methods have a number of favorable properties \cite{stats}. For example they are \textit{consistent}, i.e. they recover the true model parameters in the large sample limit. The likelihood function $\mathcal{L}$ is defined as the probability to observe data $\boldsymbol{x}$ given a model or set of parameters $\boldsymbol{\theta}$, i.e. $\mathcal{L}(\boldsymbol{\theta} | \boldsymbol{x}) = f(\boldsymbol{x} | \boldsymbol{\theta})$, with pdf $f$. To derive its form, one needs to define the observable space, as well as the signal and background models. The background in the search of neutrino point sources is generated by the conventional atmospheric \cite{Fedynitch:2015zma} and diffuse astrophysical \cite{Stettner:2019tok} neutrino fluxes. For the signal, we assume a point-like source that is located at a position $\boldsymbol{d}_{src}$ in the sky defined by right ascension $\alpha_{src}$ and declination $\delta_{src}$: $\boldsymbol{d}_{src}=\left(\alpha_{src},\,\delta_{src}\right)$. It is further assumed that this source generates neutrinos following a power law spectrum $\Phi=\Phi_0 \cdot (E_\nu/E_0)^{-\gamma}$ with neutrino energy $E_\nu$, spectral index $\gamma$, and flux normalization $\Phi_0$ at some normalization energy $E_0$.  To identify a point-source we use three observables: the estimated muon energy $\hat{E}_{\mu}$, the reconstructed muon direction $\boldsymbol{\hat{d}}$ and its estimated uncertainty $\hat{\sigma}$. Hence, a single event is characterized by the observation $\boldsymbol{x_i}$=($\boldsymbol{\hat{d}}_i$, $\hat{\sigma}_i$, $\hat{E}_{\mu,\,i}$). 
\begin{figure}[t]
   \centering
    \includegraphics{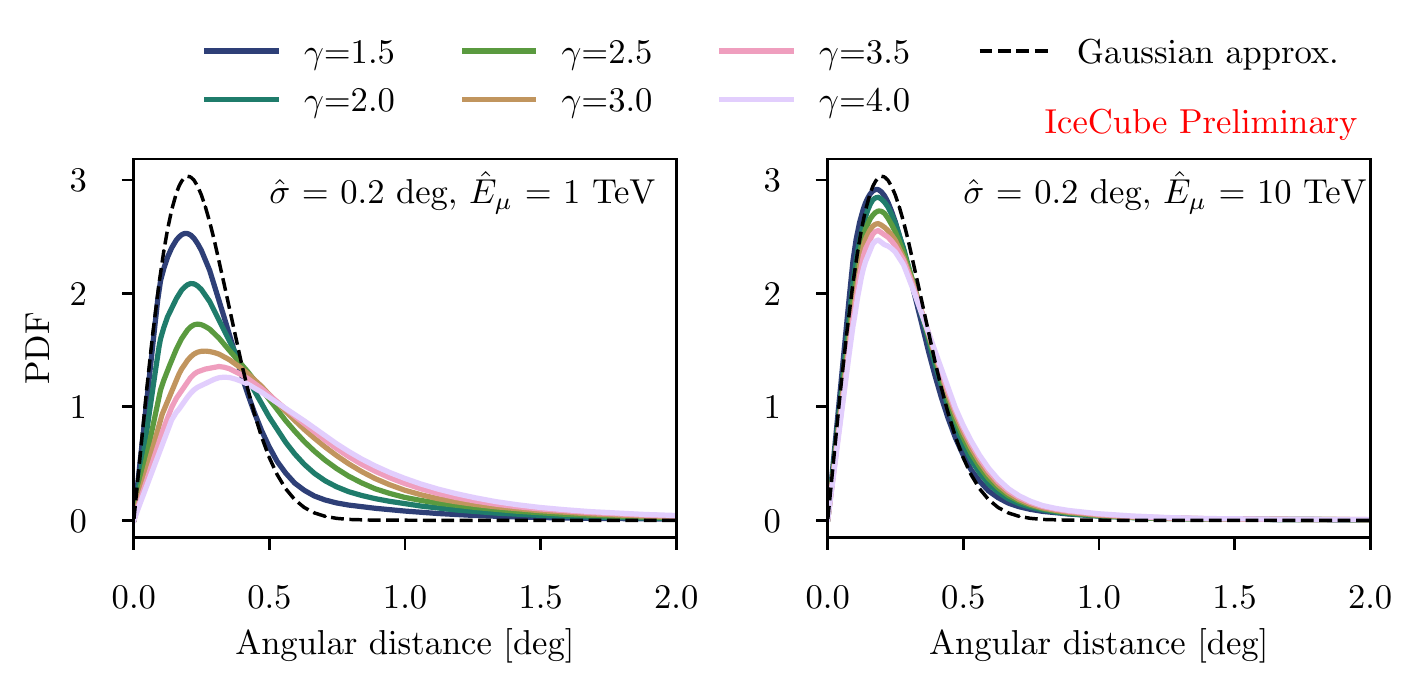}
    \caption[asd]{\textbf{Spectral index dependence of the spatial term}. Both plots show the angular distance between truth and reconstruction for an angular uncertainty $\hat{\sigma}=0.2^{\circ}$, but different energies and spectral indices. Especially in the left plot the spectral index dependence, mainly due to the kinematic angle, as well as the non-Gaussian tails are clearly visible. The Gaussian assumption, used previously, is shown as dashed line.}
    \label{fig:psf_vs_spectral_index}
\end{figure}
Due to the Earth's rotation, the \textit{background} pdf is uniform in right ascension. Thus,
\begin{equation}
\label{eq:llh_bkg}
 f_B(\boldsymbol{x_i})=f_B(\boldsymbol{\hat{d}}_i, \hat{\sigma}_i, \hat{E}_{\mu,\,i}) = \frac{1}{2\pi} f_B(\sin\hat{\delta}_i, \hat{\sigma}_i, \hat{E}_{\mu,\,i}),
\end{equation}
where $\hat{\delta}_i$ denotes the reconstructed declination. The remaining part of the pdf can be determined numerically from MC simulations. Assuming a circular reconstruction error, the \textit{source} likelihood depends only on the angular distance $\hat{\psi}_i=||\boldsymbol{\hat{d}}_i - \boldsymbol{d}_{src}||$ between reconstructed direction and source position. Hence,
\begin{equation}
f_S\left(\hat{E}_{\mu,i},\,\boldsymbol{\hat{d}_i},\,\hat{\sigma}_i\,|\,\sin\delta_{src},\,\gamma\right) = \frac{1}{2\pi \, \sin\hat{\psi}_i}\,f_S\left(\hat{E}_{\mu,i},\,\hat{\psi}_i,\,\hat{\sigma}_i\,|\,\sin\delta_{src},\,\gamma\right),
\end{equation}
where the factor $1/(2\pi\sin\hat{\psi}_i)$ ensures proper normalization on the sphere. We further separate out the spatial term $f_S(\hat{\psi}_i\,|\,\hat{\sigma}_i,\,\hat{E}_{\mu,i},\,\sin\delta_{src},\,\gamma)$ using the law of conditional probabilities 
\begin{equation}
    f_S\left(\hat{E}_{\mu,i},\boldsymbol{\hat{d}_i},\hat{\sigma}_i|\sin\delta_{src},\gamma\right) = \frac{1}{2\pi\sin\hat{\psi}_i}\,f_S\left(\hat{\psi}_i|\hat{\sigma}_i,\hat{E}_{\mu,i},\sin\delta_{src},\gamma\right) f_S\left(\hat{E}_{\mu,i},\hat{\sigma}_i|\sin\delta_{src},\gamma\right).
\end{equation}
\noindent To compromise between numerical complexity and statistical accuracy, we remove the least discriminating variable $\hat{\sigma}_i$ from the energy and background pdfs and use that, for our choice of $\hat{\sigma}_i$ (c.f. BDT sigma), the spatial term is conditionally independent of the declination $\sin\delta_{src}$, i.e.
\begin{equation}
    \label{eq:final_signal}
    f_S\left(\hat{E}_{\mu,i},\,\boldsymbol{\hat{d}_i},\,\hat{\sigma}_i\,|\,\sin\delta_{src},\,\gamma\right) \approx \frac{1}{2\pi \, \sin\hat{\psi}_i}\,f_S\left(\hat{\psi}_i\,|\,\hat{\sigma}_i,\,\hat{E}_{\mu,i},\,\gamma \right) \cdot f_S\left(\hat{E}_{\mu,i}\,|\,\sin\delta_{src},\,\gamma\right).    
\end{equation}
Previous IceCube analyses approximated the spatial term as a Gaussian that is independent of the source's spectral index $\gamma$ \cite{Braun:2008bg} and only accounts for the muon energy $\hat{E}_{\mu,i}$ through a correction $c(\hat{E}_{\mu,i})$ of the angular error, i.e., $\hat{\sigma}_i^\prime=\hat{\sigma}_i\times c(\hat{E}_{\mu,i})$,  derived assuming $\gamma=2.0$
\begin{equation}
\label{eq:approx}
f_S\left(\hat{E}_{\mu,i},\,\boldsymbol{\hat{d}_i},\,\hat{\sigma}_i^\prime\,|\,\sin\delta_{src},\,\gamma\right) \approx \frac{1}{2\pi\hat{\sigma}_i^{\prime 2}}\exp{\left(-\frac{\hat{\psi_i}^2}{2\hat{\sigma}_i^{\prime 2}}\right)} \cdot f_S\left(\hat{E}_{\mu,i}\,|\,\sin\delta_{src},\,\gamma \right).
\end{equation}
However, Figure \ref{fig:psf_vs_spectral_index} shows that those simplifications do not hold in a wide range of the observable space, in particular at low energies and small angular errors. While this does not affect the correctness of experimental p-values, a better likelihood  description improves the consistency of the method and therefore the estimation of source parameters and the potential for discoveries (c.f. Analysis Performance). In this work we extract the pdfs in eq. \eqref{eq:llh_bkg} and eq. \eqref{eq:final_signal} from MC simulations using numerical techniques \cite{Poluektov:2014rxa} \cite{Whitehorn:2013nh} and thus fully account for spectral index dependent non-Gaussian tail behavior of the spatial term.
\begin{figure}[t]
   \centering
    \includegraphics[width=0.9\textwidth]{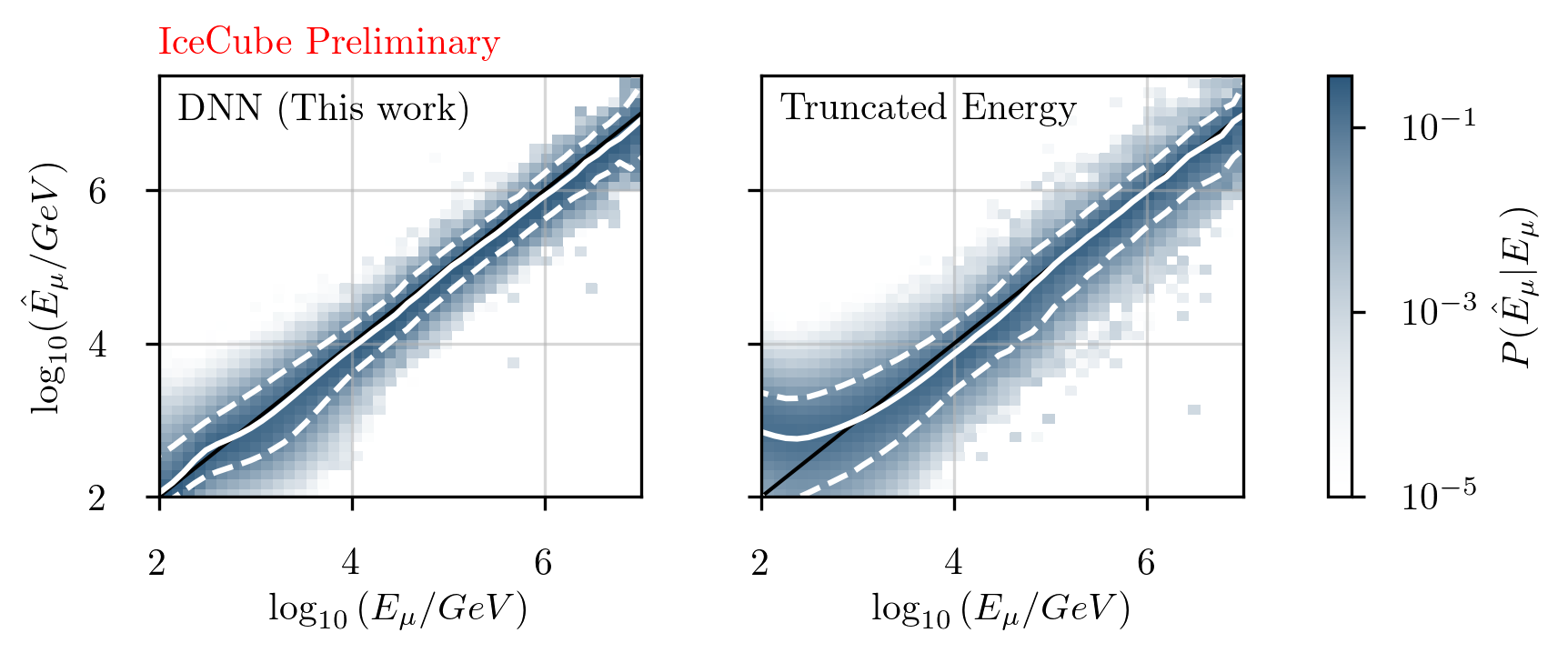}
    \caption[asd]{\textbf{Performance of the DNN energy reconstruction.} For each true energy on the x-axis, the 68\% central quantile of reconstructed energies $\hat{E}_{\mu}$ is shown as dashed and the median as white solid line for the DNN as used in the new point source analysis (left) and truncated energy, a traditional likelihood-based algorithm (right). The unbiased expectation is shown as black line.}
    \label{fig:ereco_comparison}
\end{figure}
These new numerical pdfs are then used within the likelihood function of the entire sample of $N$ events that has been described before \cite{Braun:2008bg}
\begin{equation}
    \mathcal{L}\left(\mu_{ns},\,\gamma,\,\boldsymbol{d}_{src}\,|\,\boldsymbol{x},\,N\right)=\prod_{i=1}^N\left[\frac{\mu_{ns}}{N}f_S\left(\boldsymbol{x}_i\,|\,\gamma,\,\boldsymbol{d}_{src}\right)+\left(1-\frac{\mu_{ns}}{N}\right)f_B\left(\boldsymbol{x}_i\right)\right],
\end{equation}
with observable vector $\boldsymbol{x_i}=(\boldsymbol{\hat{d}}_i$, $\hat{\sigma}_i$, $\hat{E}_{\mu,\,i})$. At any location $\boldsymbol{d}_{src}$, the test-statistic becomes
\begin{equation}
    \label{eq:ts}
    \mathcal{TS}(\boldsymbol{d}_{src}) = -2 \times \log\left(\frac{\mathcal{L}(\mu_{ns}=0\,|\, \boldsymbol{x})}{\sup_{\mu_{ns},\,\gamma}\mathcal{L}(\mu_{ns},\,\gamma\,|\,\boldsymbol{x})}\right).
\end{equation}
Its declination-dependent sampling distribution is estimated using MC simulations.\\

\noindent\textbf{New Observable: DNN Energy}

\noindent Deep neural networks (DNNs) have found wide spread application in many areas of science. In IceCube, 4-D convolutional NNs have been used for reconstruction \cite{Abbasi:2021ryj} and classification \cite{Kronmueller:2019jzh} tasks. Here, we are specifically interested in the energy at detector entry of up-going ($\delta\geq -5^{\circ}$) tracks as a proxy variable for the neutrino energy. By exchanging the output variable, we have trained a DNN with a similar Inception-ResNet architecture and training strategy as described in \cite{Kronmueller:2019jzh}. First, the IceCube In-Ice Array is transformed onto an input grid with $10 \times 10 \times 60$ pixels, including zero-padding. The pulses recorded at each digital optical module (DOM) are then converted into 15 features that carry information about the time and charge of the recorded photo-electrons, e.g. the charge recorded after $50\,\mathrm{ns}$ or the time until 10\% of the charge have been deposited. Batches of simulated muon tracks are propagated through the network and the resulting predictions are compared to the MC truth using the mean-squared error loss. To stabilise the DNN against broken DOMs in the experimental data, up to 75 DOMs and 2 strings are randomly removed during the training. The performance of the final neural network is shown in Figure \ref{fig:ereco_comparison} (left). The DNN energy estimation outperforms the traditional reconstruction ("truncated energy") (right) in resolution throughout the entire energy range. This is partly caused by having more timing information than the traditional approaches, which only takes into account the total charge at each DOM \cite{Aartsen:2013vja}. Furthermore, the DNN predictions are unbiased below 1 TeV. At these energies, continuous, energy-independent ionization starts to dominate over radiative losses. Hence, the linear relationship between the muon energy and it's energy losses breaks down. In addition, the length of the muon tracks drops below the size of the detector ($\sim$ 1 km) and decreases linearly with energy. Hence, the track length provides additional information that can be used by the DNN. The DNN estimator improves the energy resolution by up to 40\% in $\log_{10}(\hat{E}_{\mu})$ at all muon energies.\\

\noindent\textbf{New Observable: BDT Angular Error}

\noindent The precision with which IceCube reconstructs the direction of an individual muon track depends on the event properties, for example its energy and trajectory through the detector. An initial estimate of each track's angular uncertainty is provided by the directional reconstruction algorithm. While it is able to distinguish well-reconstructed tracks from poorly reconstructed ones, it does not provide correct coverage and does not account for the inherent uncertainty due to the muon's kinematic angle. Therefore, an approximate MC based calibration becomes necessary. It is typically performed in discrete bins of the reconstructed energy and assuming a spectral index of $\gamma=2$. In this work, the numerical construction of the spatial pdfs as a function of the source's spectral index guarantees correct calibration, and thus this step is not required. But, by including additional information about each event, one can still improve upon the initial angular uncertainty estimate. Here, this is done using Boosted Decision Trees (BDT) \cite{NIPS2017_6449f44a} to parameterize the median angular separation between the true direction of the track and its reconstructed value as a function of $17$ observables: total energy and stochasticity of the track's energy losses, the position in the detector where most light was deposited, the declination of the track, it's initial angular error estimate, and others. By doing so, we better account for the varying detector response to the quite diverse set of possible track signals. For example, we verified that the spatial pdf, constructed based on this new BDT angular error estimator, is independent of the track declination as assumed by the likelihood function \eqref{eq:final_signal}. This was not the case using the previous angular error estimator and likelihood methods at TeV energies and below.\\

\begin{figure}[t]
   \centering
    \includegraphics[width=0.9\textwidth]{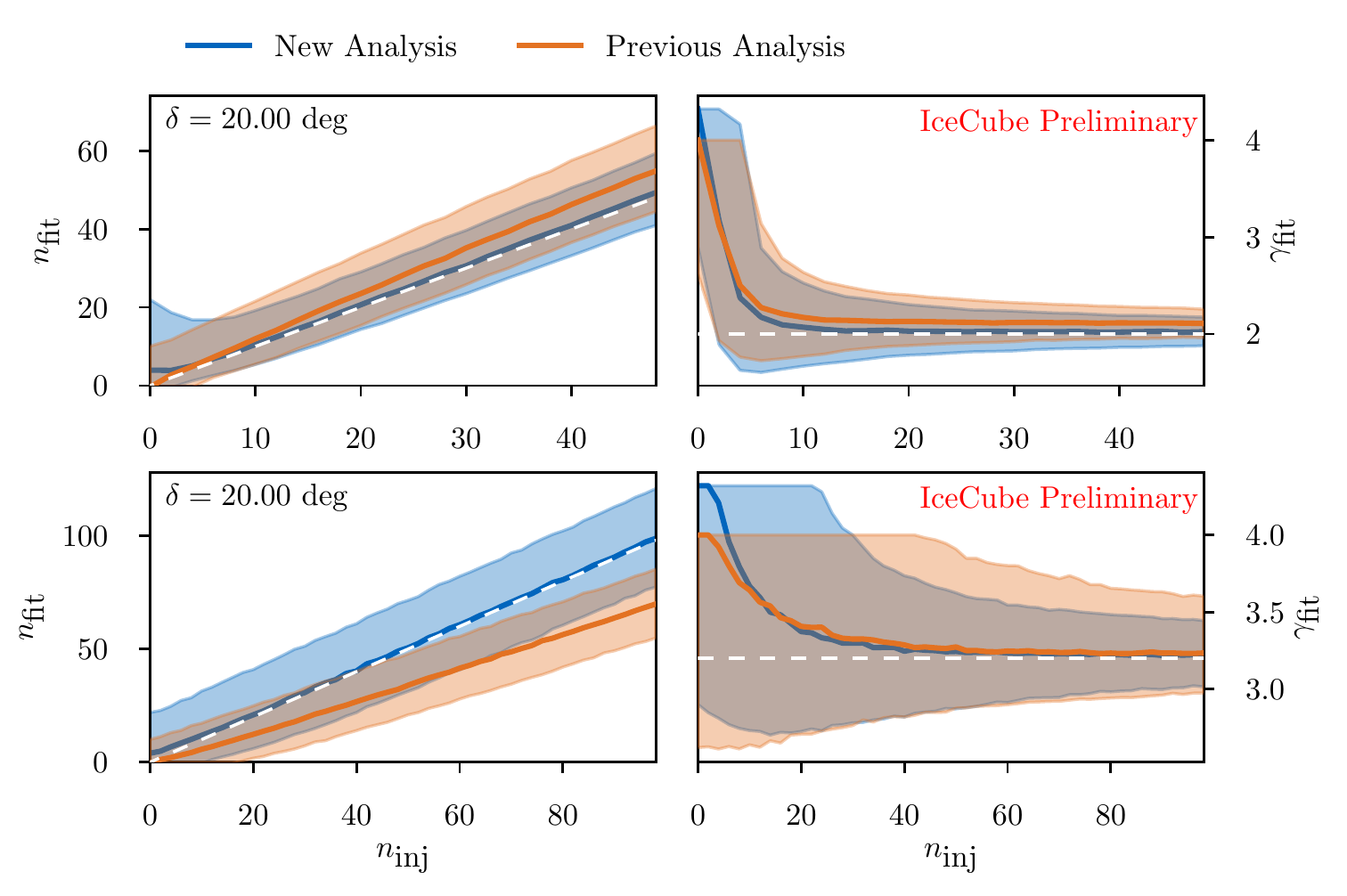}
    \caption[asd]{\textbf{Recovery of source parameters as a function of the signal strength} at a declination of $\delta=20^{\circ}$ and for two spectra $\gamma=2.0$ (top row) and $\gamma=3.2$ (bottom row). Number of signal events (left) and spectral index (right). The unbiased expectation is shown as a white dashed line. Median values for the new (blue) and the previous analysis \cite{Aartsen:2019fau} (orange) are shown as solid lines. Shaded areas show 68\% central quantiles. }
    \label{fig:ns_gamma_bias}
\end{figure}

\begin{figure}[t]
   \centering
    \includegraphics{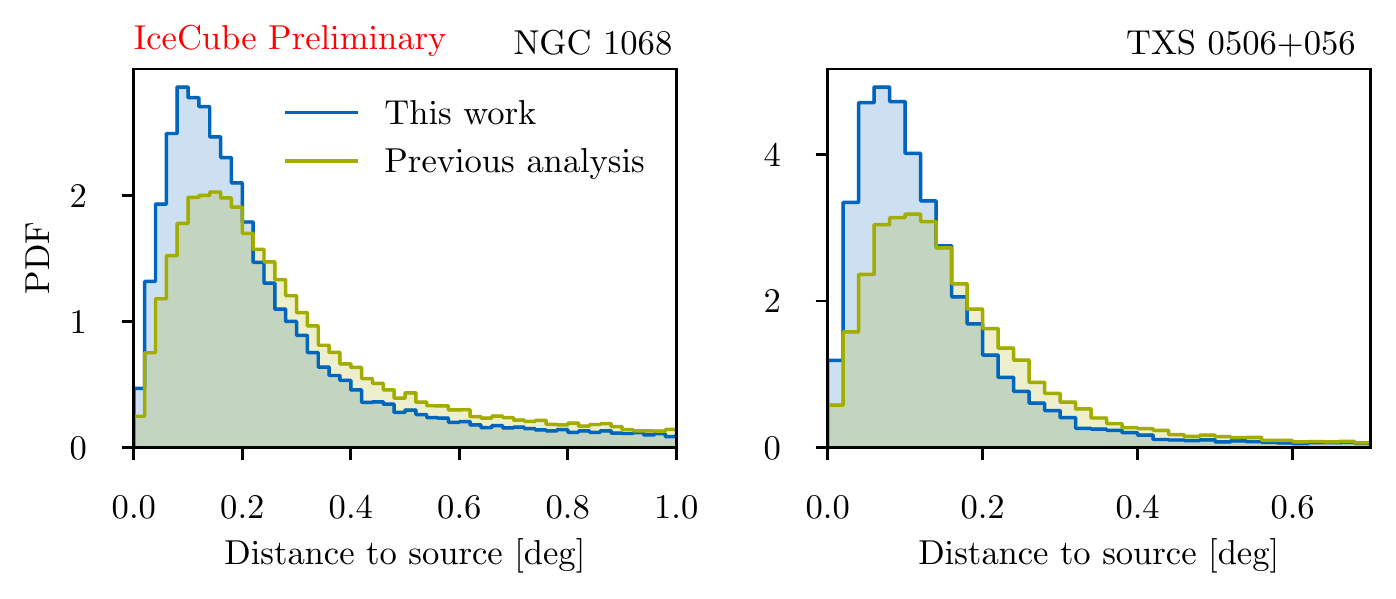}
    \caption{\textbf{Distribution of the angular separation between best-fit and true source position for this work and the previous analysis.}  Plots are based on MC simulations for the objects NGC 1068 ($\delta=-0.01^{\circ}$, $\gamma$=3.2) and TXS 0506+056 ($\delta=5.70^{\circ}$, $\gamma$=2.1) at their respective sky positions and best-fit spectra \cite{Aartsen:2019fau}.}
    \label{fig:dist_source}
\end{figure}

\noindent\textbf{Analysis Performance}\\
\noindent
The performance of the new analysis is evaluated and compared to the previous analysis \cite{Aartsen:2019fau} by generating pseudo-data with simulated point sources of varying strength and spectral index, see Figure \ref{fig:ns_gamma_bias}. 
Both, spectral index and number of signal events, are well recovered when sufficient signal is present. Hence, the new analysis is less biased to softer spectra and lower number of signal events for the $\gamma=2.0$ and $\gamma=3.2$ cases, respectively. Both problems were related to the overly-peaked shape of the spatial term, assumed Gaussian, in the likelihood.
Background events falling close to the sources' location were over-weighted ($\gamma=2.0$), and low-energy signal events located further from the source were not recovered ($\gamma=3.2$). In combination with the new observables, especially the improved energy resolution, we also observe that the variance of the estimated spectral index decreases. The recovery of signal with the new analysis works equally well at all declinations, while the biases in the previous studies tend to grow from the horizon towards the pole. The improved signal detection also leads to a more accurate localization of sources as shown in Figure \ref{fig:dist_source} for the coordinates of NGC 1068 and TXS 0506+056 with respective best-fit spectra from \cite{Aartsen:2019fau}. The median source offset of the best-fit position for the new (old) analysis are 0.24$^{\circ}$ (0.35$^{\circ}$) for NGC 1068 and 0.13$^{\circ}$ (0.21$^{\circ}$) for TXS 0506+056.
\begin{figure}[t]
   \centering
   \includegraphics[width=0.9\textwidth]{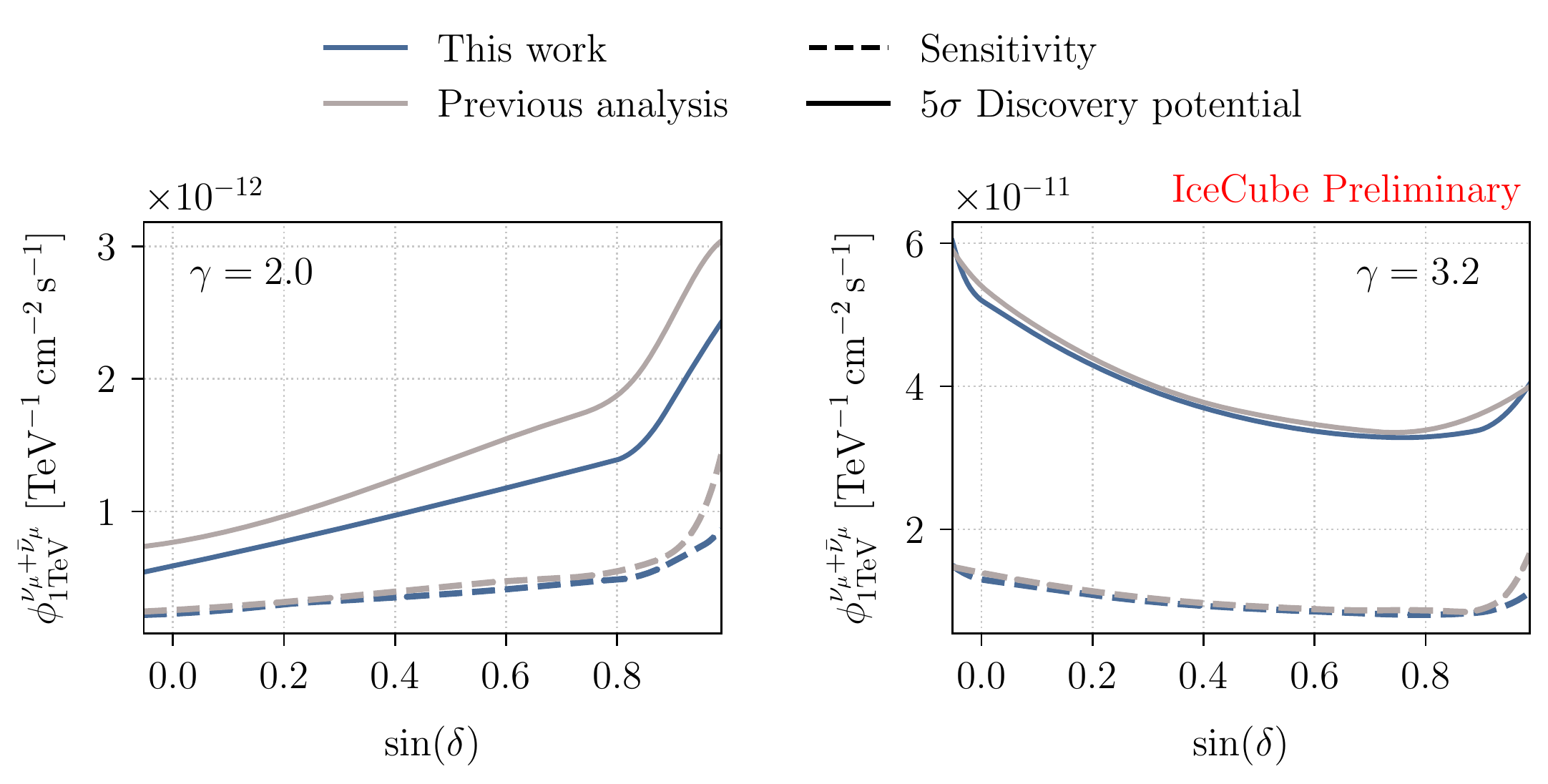}
    \caption[asd]{\textbf{Sensitivity and discovery potential as a function of source declination}. The left panel shows a hard spectrum of $\gamma=2.0$ and right a soft spectrum of $\gamma=3.2$. Different colors refer to the new and the previous analysis \cite{Aartsen:2019fau}, respectively.  }
    \label{fig:sens_dp}
\end{figure}

\noindent 
Finally, the discovery potential, i.e. the flux needed to obtain a TS value above the $5\sigma$ percentile of the background TS distribution with 50\% probability, improves with respect to the previous analysis \cite{Aartsen:2019fau}, see Figure \ref{fig:sens_dp}. The improvement is 20\% - 30\% at $\gamma=2.0$ and 5\% at $\gamma=3.2$. Sensitivities remain basically the same as the improvements become relevant at high signal strength.\\ 

\noindent\textbf{Summary and Outlook}

\noindent We have presented a novel point source analysis framework for the IceCube data, that introduces several improvements. We have implemented a more accurate description of the likelihood function through relaxation of previous assumptions. The spatial clustering of events around a point-source is described more correctly, accounting also for its dependence on the source's spectral index for any combination of energy and angular error estimates. A new DNN-based energy reconstruction improves resolution and resolves degeneracies at muon energies below a few TeV. These changes lead to an improved discovery potential by up to 30\% ($\gamma=2.0$) and, probably even more important, enable an unbiased estimation of the source parameters. 
Also, the localization uncertainty of a possible point-source signal is reduced. 
IceCube has recently reprocessed all of its experimental data since 2010, applying the latest detector calibration as well as unified filtering. These changes also improve the accuracy of event directions, allowing for significance increases beyond what is expected from purely methodological changes. The results of a search for point sources in 9 years of data from the full detector configuration including the new likelihood, improved reconstructions, and reprocessed data are currently under final review and are being prepared for publication.

\bibliographystyle{ICRC}
\bibliography{references}



\clearpage
\section*{Full Author List: IceCube Collaboration}




\scriptsize
\noindent
R. Abbasi$^{17}$,
M. Ackermann$^{59}$,
J. Adams$^{18}$,
J. A. Aguilar$^{12}$,
M. Ahlers$^{22}$,
M. Ahrens$^{50}$,
C. Alispach$^{28}$,
A. A. Alves Jr.$^{31}$,
N. M. Amin$^{42}$,
R. An$^{14}$
% ...
\\

\noindent
% ...
$^{12}$ Universit{\'e} Libre de Bruxelles, Science Faculty CP230, B-1050 Brussels, Belgium \\
% ...
$^{14}$ Department of Physics and Laboratory for Particle Physics and Cosmology, Harvard University, Cambridge, MA 02138, USA \\
% ...
$^{17}$ Department of Physics, Loyola University Chicago, Chicago, IL 60660, USA \\
$^{18}$ Dept. of Physics and Astronomy, University of Canterbury, Private Bag 4800, Christchurch, New Zealand \\
% ...
$^{22}$ Niels Bohr Institute, University of Copenhagen, DK-2100 Copenhagen, Denmark \\
% ...
$^{28}$ D{\'e}partement de physique nucl{\'e}aire et corpusculaire, Universit{\'e} de Gen{\`e}ve, CH-1211 Gen{\`e}ve, Switzerland \\
% ...
$^{31}$ Karlsruhe Institute of Technology, Institute for Astroparticle Physics, D-76021 Karlsruhe, Germany  \\
% ...
$^{42}$ Bartol Research Institute and Dept. of Physics and Astronomy, University of Delaware, Newark, DE 19716, USA \\
% ...
$^{50}$ Oskar Klein Centre and Dept. of Physics, Stockholm University, SE-10691 Stockholm, Sweden \\
% ...
$^{59}$ DESY, D-15738 Zeuthen, Germany \\
% ...

\subsection*{Acknowledgements}

\noindent
USA {\textendash} U.S. National Science Foundation-Office of Polar Programs,
U.S. National Science Foundation-Physics Division,
U.S. National Science Foundation-EPSCoR,
Wisconsin Alumni Research Foundation,
Center for High Throughput Computing (CHTC) at the University of Wisconsin{\textendash}Madison,
Open Science Grid (OSG),
Extreme Science and Engineering Discovery Environment (XSEDE),
Frontera computing project at the Texas Advanced Computing Center,
U.S. Department of Energy-National Energy Research Scientific Computing Center,
Particle astrophysics research computing center at the University of Maryland,
Institute for Cyber-Enabled Research at Michigan State University,
and Astroparticle physics computational facility at Marquette University;
Belgium {\textendash} Funds for Scientific Research (FRS-FNRS and FWO),
FWO Odysseus and Big Science programmes,
and Belgian Federal Science Policy Office (Belspo);
Germany {\textendash} Bundesministerium f{\"u}r Bildung und Forschung (BMBF),
Deutsche Forschungsgemeinschaft (DFG),
Helmholtz Alliance for Astroparticle Physics (HAP),
Initiative and Networking Fund of the Helmholtz Association,
Deutsches Elektronen Synchrotron (DESY),
and High Performance Computing cluster of the RWTH Aachen;
Sweden {\textendash} Swedish Research Council,
Swedish Polar Research Secretariat,
Swedish National Infrastructure for Computing (SNIC),
and Knut and Alice Wallenberg Foundation;
Australia {\textendash} Australian Research Council;
Canada {\textendash} Natural Sciences and Engineering Research Council of Canada,
Calcul Qu{\'e}bec, Compute Ontario, Canada Foundation for Innovation, WestGrid, and Compute Canada;
Denmark {\textendash} Villum Fonden and Carlsberg Foundation;
New Zealand {\textendash} Marsden Fund;
Japan {\textendash} Japan Society for Promotion of Science (JSPS)
and Institute for Global Prominent Research (IGPR) of Chiba University;
Korea {\textendash} National Research Foundation of Korea (NRF);
Switzerland {\textendash} Swiss National Science Foundation (SNSF);
United Kingdom {\textendash} Department of Physics, University of Oxford.

\end{document}